\documentclass[aps,prb,twocolumn,floatfix,preprintnumbers,superscriptaddress,amsmath,amssymb]{revtex4-1}

\usepackage{xcolor,bm,lineno,multirow,float,times}
\usepackage{physics}
\usepackage{rotating}
\usepackage{verbatim}
\usepackage{graphicx}
\usepackage[multiple]{footmisc}
\usepackage{natbib}
\usepackage{soul}

\usepackage[%
  colorlinks=true,
  urlcolor=blue,
  linkcolor=blue,
  citecolor=blue
]{hyperref}

\newcommand{\blue}[1]{{\color{black} #1}}

\def\navg{$n_{\mathrm{p}}$}

\begin{document}

\title{Neutron scattering investigation of proposed Kosterlitz-Thouless transitions in the triangular-lattice Ising antiferromagnet TmMgGaO${_{4}}$}
	
\author{Zhiling Dun}
\affiliation{School of Physics, Georgia Institute of Technology, Atlanta, GA 30332, USA}

\author{Marcus Daum}
\affiliation{School of Physics, Georgia Institute of Technology, Atlanta, GA 30332, USA}
	
\author{Raju Baral}
\affiliation{Department of Physics and Astronomy, Brigham Young University, Provo, UT 84602, USA}

\author{Henry E. Fischer}
\affiliation{Institut Laue-Langevin, BP 156, 38042 Grenoble Cedex 9, France}

\author{Huibo  Cao}
\affiliation{Neutron Scattering Division, Oak Ridge National Laboratory, Oak Ridge, TN 37831, USA}

\author{Yaohua Liu}
\affiliation{Neutron Scattering Division, Oak Ridge National Laboratory, Oak Ridge, TN 37831, USA}

\author{Matthew B. Stone}
\affiliation{Neutron Scattering Division, Oak Ridge National Laboratory, Oak Ridge, TN 37831, USA}

\author{Jose A. Rodriguez-Rivera}
\affiliation{Department of Materials Sciences, University of Maryland, College Park, Maryland 20742, USA}
\affiliation{NIST Center for Neutron Research, Gaithersburg, MD 20899, USA}

\author{Eun Sang Choi}
\affiliation{National High Magnetic
Field Laboratory, Florida State University, Tallahassee, FL, 32306, USA}

\author{Qing Huang}
\affiliation{Department of Physics and Astronomy, University of Tennessee, Knoxville, TN 37996, USA}

\author{Haidong Zhou}
\affiliation{Department of Physics and Astronomy, University of Tennessee, Knoxville, TN 37996, USA}

\author{Martin Mourigal}
\affiliation{School of Physics, Georgia Institute of Technology, Atlanta, GA 30332, USA}

\author{Benjamin A. Frandsen}
\affiliation{Department of Physics and Astronomy, Brigham Young University, Provo, UT 84602, USA}

\date{\today}

\begin{abstract}
The transverse-field Ising model on the triangular lattice is expected to host an intermediate  finite-temperature Kosterlitz-Thouless (KT) phase through a mapping of the spins on each triangular unit to a complex order parameter. TmMgGaO$_4$ is a candidate material to realize such physics due to the non-Kramers nature of the Tm$^{3+}$ ion and the resulting two-singlet single-ion ground state. Using inelastic neutron scattering, we confirm this picture by determining the leading parameters of the low-energy effective Hamiltonian of TmMgGaO$_4$. Subsequently, we track the predicted KT phase and related transitions by inspecting the field and temperature dependence of the ac susceptibility. We further probe the spin correlations in both reciprocal space and real space via single crystal neutron diffraction and magnetic total scattering techniques, respectively. Magnetic pair distribution function analysis provides evidence for the formation of vortex-antivortex pairs that characterize the proposed KT phase around 5~K. Although structural disorder influences the field-induced behavior of TmMgGaO$_4$, the magnetism in zero field appears relatively free from these effects. These results position TmMgGaO$_4$ as a strong candidate for a solid-state realization of KT physics in a dense spin system.
\end{abstract} 

\maketitle

\section{Introduction}
Interacting Ising spins in a transverse magnetic field  display important quantum many-body effects \cite{Stinchcombe1973}, including quantum phase transitions \cite{Ronnow2005,Suzuki2012} and order by disorder \cite{Moessner2000a}. Frustrated magnets comprised of non-Kramers ions in a low-symmetry crystal field, such as the pyrochlore materials Pr${_2}$Zr(Hf)${_2}$O${_7}$ \cite{kimura2013,Wen2017,Sibille2018} and the kagome magnet Ho${_3}$Mg${_2}$Sb${_3}$O${_{14}}$ \cite{Dun2017,Dun2019}, host a two-singlet ground-state that maps onto an intrinsic transverse field acting on Ising spins~\cite{Wang1968}, which can promote quantum fluctuations \cite{Benton2017} and entanglement \cite{Savary2017}. In models where such magnetic ions decorate a triangular lattice, the transverse field induces a three-sublattice (3SL) order for antiferromagnetically coupled Ising spins through a quantum order by disorder phenomenon \cite{Villain1980, Moessner2000a}. This can be re-cast as a two-dimensional $XY$ model with a $Z_6$ clock term \cite{Moessner2000, Damle2015}, for which two finite-temperature Kosterlitz-Thouless (KT) transitions are expected to border an intermediate phase with power-law spin correlations~\cite{Jose1977}. These deep theoretical insights offer the enticing opportunity to realize the topological vortex-pair binding and unbinding transitions proposed by Kosterlitz and Thouless in 1973 \cite{Kosterlit1973} in a dense spin system.

\begin{figure}[tbp!]
	\linespread{1}
	\par
	\begin{center}
		\includegraphics[width= \columnwidth ]{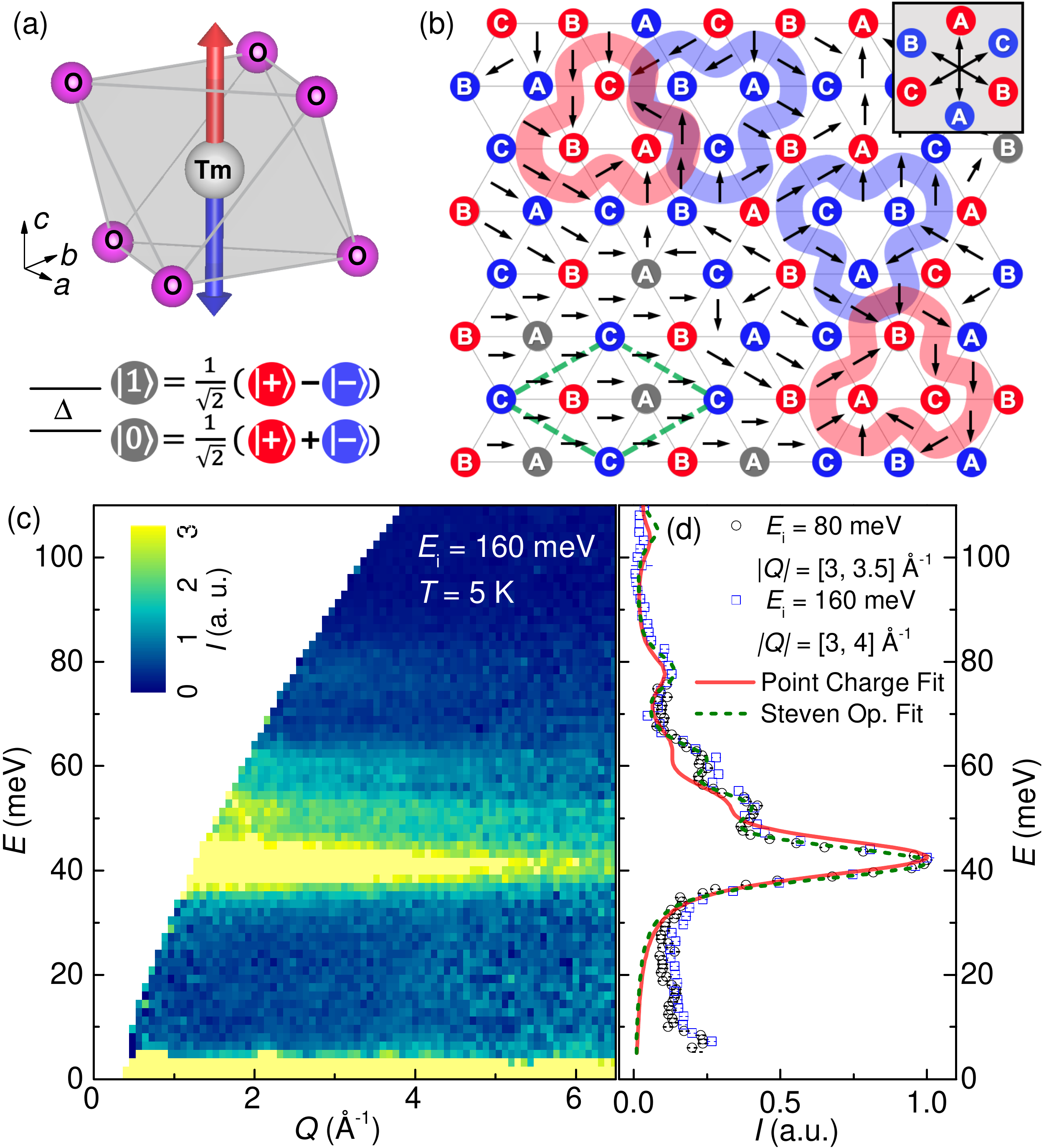}
	\end{center}
	\par
	\caption{\label{Fig:CEF} (a) Oxygen ligands (pink spheres) around each Tm$^{3+}$ ion (grey sphere). The two-singlet crystal electric field ground state comprises $\ket{0}$  and $\ket{1}$, which are symmetric and antisymmetric combinations of spin up ($\ket{+}$, red arrow) and spin down ($\ket{-}$, blue arrow). (b) Various spin and pseudospin configurations in TmMgGaO$_4$. Red, blue, and grey circles represent up ($\ket{+}$), down ($\ket{-}$), and nonmagnetic ($\ket{0}$) spins, respectively, with the labels A, B, and C corresponding to the three sublattices. The 3SL order is displayed in the lower left area of the diagram, with the dashed green lines showing the magnetic unit cell. The  black arrows represent the pseudospin $\psi$ defined in Eq.~\ref{eq:psi}, with the top right grey panel schematically illustrating the complex phase factors applied in the equation. The thick red (blue) shaded paths show vortices (antivortices) in the spatial arrangement of $\psi$, characterized by an emergent topological charge equal to the winding number of $\psi$ around the path \cite{Li2020nc}. The overlapping vortices and antivortices show the two possible types of bound V-AV pairs considered in this work. (c) Neutron scattering intensity $I(Q,E)$ from broad-band measurements of TmMgGaO$_4$ at $T$ = 5\,K and $E_i=$ 160 meV. (d) Scattering intensity at low momentum-transfer $Q$. The intensity was integrated within given $\abs{Q}$ ranges and normalized by the maximum intensity. Two data sets measured with different incident energies of 80 meV (blue squares) and 160 meV (black circles) are shown. The solid red curve and dashed green curve represent the best fit to the spectra using the effective point-charge approach ~\cite{Dun2020} and the Stevens operator approach, respectively.}
\end{figure}

The recently synthesized rare-earth antiferromagnet TmMgGaO$_4$~\cite{CEVALLOS2018} has been proposed to realize the transverse-field Ising model on the triangular-lattice \cite{Shen2019, Li2020}. This material derives from the intensely studied quantum spin-liquid candidate YbMgGaO$_4$ \cite{Li2015, Shen2016, Paddison2017}, where Yb$^{3+}$, a Kramers ion, is replaced by Tm$^{3+}$, a non-Kramers ion. In TmMgGaO$_4$, the crystal electric field forces the dipolar magnetic moments to point out of the triangular plane (crystallographic $c$-axis) while the in-plane components transform as magnetic multipoles whose correlations are not directly observable by x-ray or neutron scattering techniques \cite{Shen2019,Liu2020}. The two-singlet ground-state, and thus the transverse field, appears accidentally in TmMgGaO$_4$ from the octahedral environment of oxygen ligands [Fig. \ref{Fig:CEF}(a)]~\cite{Dun2020}. 

Previous neutron scattering studies of TmMgGaO$_4$ \cite{Shen2019, Li2020} uncovered the predicted 3SL order below $T=$ 1~K, although no corresponding anomalies were observed in specific heat or magnetic susceptibility measurements. Theoretical studies suggest that the 3SL order corresponds to the low-temperature transition ($T\!\equiv\!T_l$) out of the proposed KT phase \cite{Li2020nc, Liu2020}, and that there should exist another KT transition at a higher temperature ($T_h\!\approx\!$ 4\,K) \cite{Li2020nc} corresponding to the unbinding of vortex-antivortex (V-AV) pairs. These vortices emerge via a mapping from the $S^z$ components of the Tm$^{3+}$ spins to a complex order parameter (or pseudospin) $\psi$ given by
\begin{equation}
\label{eq:psi}
\psi= \abs{\psi}\mathrm{e}^{i\theta}=S_{\mathrm{A}}^{z}+\mathrm{e}^{i 2 \pi / 3} S_{\mathrm{B}}^{z}+\mathrm{e}^{i 4 \pi / 3} S_{\mathrm{C}}^{z},
\end{equation} 
where A, B, and C are the sublattice indices of the 3SL order \cite{Wang2017, Li2020nc}. Through this mapping [illustrated in Fig. ~\ref{Fig:CEF}(b)],  the 3SL order corresponds to ferromagnetic ordering of $\psi$ and cannot lead to any pseudospin vortices. However, vortices \textit{can} originate from short-range correlations between physical Tm$^{3+}$ spins and/or the presence of defects in otherwise long-range ordered Tm$^{3+}$ spin configurations~\cite{Li2020nc}. In the proposed scenario for TmMgGaO$_4$, then, the long-range 3SL order melts at $T_l$, leading to short-range Tm$^{3+}$ spin correlations above $T_l$ that host bound pseudospin V-AV pairs. As the temperature is raised further, the vortices and antivortices eventually unbind at $T_h$~\cite{Li2020nc}. Thus, the KT phase exists for $T_l < T < T_h$. To date, very few experimental signatures of the proposed KT transition at $T_h$ have been reported, although a recent NMR study found a hump in the 1/$T$ dynamics between 0.9\,K and 1.9\,K that was interpreted as evidence for $T_h$. Moreover, in light of the intrinsic Mg-Ga structural disorder that strongly affects the low-temperature magnetism of the isostructural compound YbMgGaO$_4$ \cite{Paddison2017, Li2017, Zhu2017, Kimchi2018,Li2020}, it is unclear whether or not KT physics should even be expected to survive in TmMgGaO$_4$.  For instance, it was recently proposed that weak quenched nonmagnetic disorder could drive the emergent KT phase into a gauge glass phase instead\cite{huang2020emergent}.

In this work, we present a series of magnetometry and neutron scattering measurements that provide evidence for the proposed KT transitions in TmMgGaO$_4$ at $T_h\approx 5$\,K and $T_l \approx 0.9$\,K. We use ac magnetometry to investigate scaling predictions in the KT regime \cite{Damle2015, Biswas2018, Li2020nc} and neutron scattering measurements in the field-polarized and paramagnetic phases to ascertain the material's Hamiltonian. We then track the temperature-dependent Tm$^{3+}$ spin correlations in both reciprocal space and real space, the latter using magnetic pair distribution function (mPDF) analysis~\cite{frand;aca14,frand;aca15}. Our analysis reveals a continuous increase in correlation length over almost two decades in temperature, with experimental signatures for two possible transitions at $T_h\approx 5$\,K and $T_l \approx 0.9$\,K.  We find that $T_l$ corresponds to the gradual condensation of two-dimensional magnetic scattering at the $K$-point of the triangular Brillouin zone (corresponding to the 3SL order), although the correlation length remains finite in the plane down to at least 50\, mK. The mPDF analysis reveals short-range spin correlations that are consistent with the formation of bound V-AV pairs and which show the expected temperature dependence in and above the KT phase. Taken together, these results provide solid -- albeit indirect -- evidence for KT physics in TmMgGaO$_4$ and reveal that the significant structural disorder in this compound does not appear to profoundly impact the zero-field physics.

\section{Experimental Methods}
Single-crystal samples of TmMgGaO$_4$ were synthesized using a floating zone furnace \cite{CEVALLOS2018}.  A small crystal  with a natural cleft along the crystallographic $c$-axis was used for the ac susceptibility ($\chi_{ac}$) measurements.  Measurements above  and below 1.7\,K  were performed using a commercial Quantum Design Physical Property Measurement (PPMS) system and a home-built apparatus at SCM2 of National High Magnetic Field Laboratory \cite{Dun2014}, respectively. The  high temperature part of the SCM2 data was scaled to the PPMS data to convert the  data set into absolute units of emu/mol/Oe. With the ac field applied along the crystallographic $c$-axis of TmMgGaO$_4$, the measured $\chi_{ac}$ signals are independent of ac field frequency  ($f$, 80 - 1 kHz) and ac field magnitude ( $H_\mathrm{ac}$, 3.2 - 10 Oe).  The data shown in this manuscript were obtained with $H_\mathrm{ac}$ = 10 Oe and $f$ = 80 Hz.

Elastic single crystal neutron scattering measurements were carried out at the Four-Circle Diffractometer (HB3A) \cite{Chakoumakos2011} at the High Flux Isotope Reactor and the Elastic Diffuse Scattering Spectrometer (CORELLI) \cite{Ye2018} at the Spallation Neutron Source, both located at Oak Ridge National Laboratory. For the HB3A measurement, a single crystal was oriented in the $HK0$ scattering plane and polished into a disk shape, with a diameter of 4 mm and thickness of 2\,mm [see Fig.~4(c)], to minimize and correct for the neutron absorption of Tm. A constant neutron wavelength ($\lambda = 1.551$~\AA) was used throughout the experiment. For the CORELLI measurements, a single crystal was first  oriented in the $HK0$ scattering plane and cooled down to 2\,K using an orange cryostat inside a 5\,T Slim SAM magnet. Measurements were performed over 180$^{\circ}$ of sample rotation with 2$^{\circ}$ per step. An empty cryostat measurement was performed separately to serve as background. The crystal was then reoriented in the $HHL$ scattering plane and cooled down to a base temperature of 50\,mK with a dilution refrigerator.  Measurements were performed over 300$^{\circ}$ of sample rotation at $T$ = 50\,mK, 400\,mK, 800\,mK, and 40\,K. 

Inelastic neutron scattering measurements were carried out using the Fine-Resolution Fermi Chopper Spectrometer (SEQUOIA) \cite{Granroth2010} at the Spallation Neutron Source at Oak Ridge National Laboratory and using the Multi-Axis Crystal Spectrometer (MACS) \cite{Rodriguez2008} at the NIST Center for Neutron Research.  For the SEQUOIA experiment, a single crystal of $\sim$1\,g (size: 4 mm$\times$10 mm$\times$1.5 mm) was cooled to 5\,K with a closed-cycle refrigerator. Crystal electric field excitations were measured with incident neutron energies of $40$, $80$, and $160$ meV. The same measurements were repeated for an empty aluminum sample holder and used for background subtraction.  
For the MACS measurements, the same crystal was mounted onto a copper plate and cooled to $50$\,mK using a dilution refrigerator with a 6 T magnet. The measurements were first performed at zero field at selected energy transfers of 0, 0.4, 0.8, and 1.2 meV, with a fixed final neutron energy of $3.7$\,meV \cite{Supplymental}. The spin wave dispersion of the polarized state was mapped out under an external magnetic field of $\mu_0 H = 5.5 T$ applied along the crystallographic $c$-axis, with a fixed final neutron energy of $5$\,meV and at fixed energy transfer between 1.0 and 2.6\,meV with a step of \,0.1 meV. 

Neutron total scattering experiments were performed on a carefully ground single crystalline sample of TmMgGaO$_4$ using the D4 instrument at the Institut Laue-Langevin with a constant wavelength beam ($\lambda = 0.5$~\AA). The sample was loaded into a vanadium can with annular geometry and placed in a cryofurnace with a base temperature of 3~K. Energy-integrated total scattering patterns were collected at several temperatures between 3~K and 50~K and were reduced according to standard protocols at D4 to obtain the absolutely normalized total scattering structure function $S(Q)$. The total pair distribution function (PDF) $G_{\mathrm{total}}(r)$ was obtained via the Fourier transform of $Q(S(Q)-1)$ with $Q_{\mathrm{max}} = 24~\mathrm{\AA^{-1}}$. Atomic PDF fits were carried out in PDFgui~\cite{farro;jpcm07} using the published crystallographic structure. A representative fit to the data at 50~K is displayed in Supplemental Information \cite{Supplymental}. The refined parameters agree closely with published results \cite{CEVALLOS2018}. mPDF analysis was conducted using the diffpy.mpdf package in the DiffPy suite~\cite{juhas;aca15}, and Reverse Monte Carlo (RMC) modelling was done using home-built python code.

\section{Crystal electric field excitations}
\begin{figure}[tbp!]
	\linespread{1}
	\par
	\begin{center}
		\includegraphics[width= \columnwidth ]{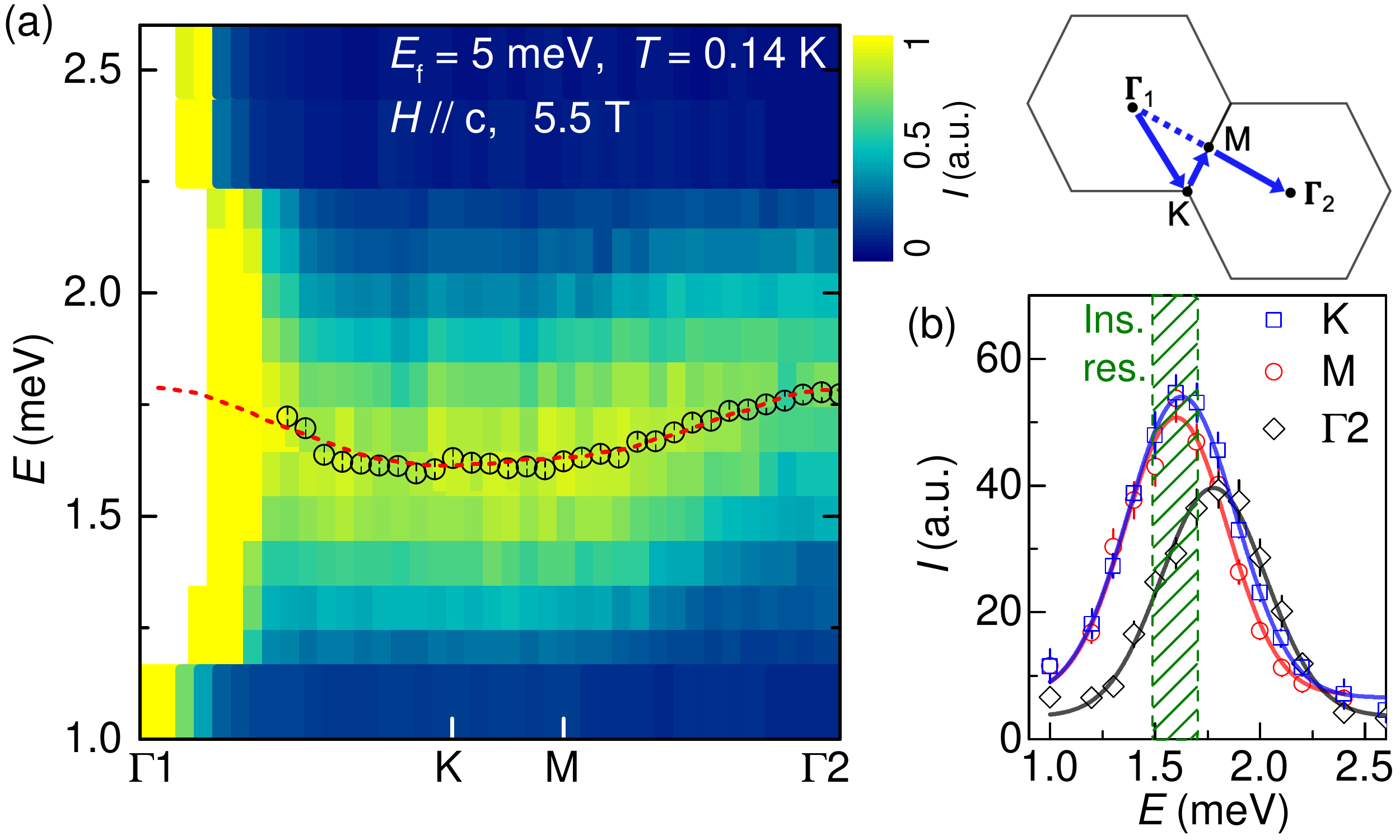}
	\end{center}
	\par
	\caption{\label{Fig:INS}  (a) Low-energy spin excitations along four high symmetric directions in reciprocal space (top right corner) at $T\!=\!50$\,mK under an external magnetic field of $\mu_0H\!=\!5.5\,T$ applied along the $c$-axis. Solid circles represent the peak centers from Gaussian fits to constant-$Q$ cuts, and the dashed red line represents the best fit using Eq. \ref{eq:TIM} based on linear spin wave theory.  (b) Constant-$Q$ cuts (open symbols) at three high-symmetric positions with Gaussian fits (solid lines). The instrument resolution is approximately $0.2$ meV and is illustrated by the dashed green area.}
\end{figure}

We start with the validation of the effective spin-1/2 Hamiltonian for TmMgGaO$_4$. Broadband inelastic neutron scattering experiments using the SEQUOIA spectrometer~\cite{Granroth2010} at Oak Ridge National Laboratory (ORNL) reveal crystal electric-field (CEF) excitations for an energy transfer $E$ between 30 and 90 meV, with the expected decrease in scattering intensity $I(Q,E)$ at large momentum transfer ${Q}$ [Fig. \ref{Fig:INS}(a)]. Excitations are observed at 41.9(1), 52.6(1), 61.4(2), and 78.7(2) meV [black circles, Fig. \ref{Fig:INS}(b)], with the 41.9 meV mode corresponding to the first excited CEF level above the two-singlet ground state. The $\Delta E\approx7.5$--$10$~meV width of the CEF peaks does not depend on the incident neutron energy, and is much broader than the instrumental resolution, suggesting the origin of this broadening is intrinsic to the material, reminiscent of YbMgGaO$_4$ \cite{Paddison2017, Li2017}. We use two methods to model the excitation spectrum: the conventional Stevens operator approach, and the effective point charge model outlined in Ref.~\cite{Dun2020} [Fig.~\ref{Fig:INS}(c)]. \blue{The former method contains six $B^{m}_{n}$  CEF parameters due to the 3-fold symmetry of Tm$^{3+}$ ion, while the latter method reduces the fit variables to three, namely a distance ($r$), an angle  ($\theta$), and an effective point charge value ($q$) \cite{Dun2020}.}  Both fits capture the energy of the four CEF excitations successfully and yield similar CEF parameters [Table ~\ref{table:CEF}]. 

The analysis from both fits yields a CEF ground state comprising two singlets, $\ket{0}$ and $\ket{1}$, which can be expressed as symmetric and antisymmetric combinations non-Kramers doublet states $\ket{+}$ and $\ket{-}$ as 
\begin{equation}
 \ket{0} \approx \frac{1}{\sqrt{2}} (\ket{+}+\ket{-}),  \ket{1}  \approx  \frac{1}{\sqrt{2}} (\ket{+}-\ket{-}).
\end{equation}
Using the numerical results from the effective point charge fits shown in Table~\ref{table:CEF}, we can express the non-Kramers doublet states in the $\ket{J=6, J_z}$ basis as
\begin{equation}
 \ket{\pm}  =0.924\ket{\pm6} + 0.339\ket{\pm3} + 0.085\ket{0},
\end{equation}
which represent spin up and down, respectively [see Fig.\ref{Fig:CEF}(a)]. 
Our CEF analysis validates the mapping to an effective transverse-field Ising model \cite{Dun2019, Shen2019} by which the CEF Hamiltonian reads $\mathcal{H}_\textrm{CF}  = \Delta S^x$, where $\Delta$ is the splitting between the two singlets, and the magnetic moments map onto spin-1/2 with an effective $g$-factor of $g_\mathrm{eff}$ = 2$g_{J}\bra{0}J_z\ket{1}$.

\blue{ 
The values of $\Delta$ determined from the Stevens operators and effective point charge approach are 0.09 and 0.28~meV, respectively. These are comparable to, and thus strongly modified by, spin-spin interactions, precluding a direct neutron-scattering measurement of this energy.  Moreover,  $\Delta$ is two orders of magnitude smaller than the observed CEF energies, which contributes to a relatively large uncertainty for the fitted value $\Delta$. 
Although the Stevens operator fit agrees with the measured data better (particularly around 50 to 60\,meV), the effective point charge fit may nevertheless provide a better estimate of  $\Delta$ and $g_\mathrm{eff}$. This is because structural disorder, which results in  a distribution of both $g_\mathrm{eff}$ and $\Delta$ \cite{Li2020}, is not taken into account in the CEF fits. With more fitting parameters, the Stevens operator approach is prone to over-fitting to features originating from disorder. In contrast, the effective point-charge fit, which directly incorporates the local crystallography, may better reflect the intrinsic nature of CEF properties of the average structure model in TmMgGaO$_4$. This is borne out by more reliable estimates of $\Delta$ inferred from the high-field spin excitations (see Section~\ref{high-field}), which indicate $\Delta = 0.6$~meV, and from magnetization measurements leading to $g_\mathrm{eff} = 13.2$.~\cite{Li2020} 
}

\begin{table*}[tbp]  
		\centering
		\caption{\label{table:CEF} Tabulated CEF parameters, CEF energies ($E^\textrm{CEF}_i$), wave-functions of the two-singlet ground state ($\ket{0}$  and $\ket{1}$) in the total angular momentum $\ket{J=6, J^z}$ basis, and effective $g$-factor ($g_\mathrm{eff}$) from the effective point charge (PC) fit \cite{Dun2020} and Stevens operator fit to the CEF excitation spectra of TmMgGaO$_4$.  For $E^\textrm{CEF}_i$, underlined values indicate singlet CEF levels.}
		\begin{tabular}{c|cccccc}
		\hline\hline 
	      \multicolumn{7}{c}{Point Charge Fit }  \\
	    \hline
		 PC parameters   & \multicolumn{6}{c}{$r = 1.643,\mathrm{\AA},  \;\;  \theta = 59.83 ^\circ , \;\;  q = 0.519\,e$} \\
		 \hline
		 CEF parameters    & $B_2^0 $ & $B_4^0$ & $B_4^3 $ & $B_6^0  $ & $B_6^3 $ & $B_6^6$  \\
		    (meV)          &  -0.518 & -3.13e-3 & 0.120 & -3.28e-5  & 8.31e-5 & -3.06e-4 \\
		 \hline
		 $E^\textrm{CEF}_i$ (meV) & \multicolumn{6}{c}{ \underline{0}, \underline{0.28},  41.8, 51.8,  \underline{63.5},  \underline{76.7}, 107.0, \underline{116.8} ,116.4}\\
		\hline
		 $\ket{0}$   & \multicolumn{6}{l}{   $0.654(\ket{6}+\ket{-6})-0.240(\ket{3}-\ket{-3})+ 0.170\ket{0}$}\\
	      $\ket{1}$   & \multicolumn{6}{l}{   $0.659(\ket{6}-\ket{-6})-0.257(\ket{3}+\ket{-3})$}\\
	      \hline
	      $g_\mathrm{eff}$   & \multicolumn{6}{c}{ 13.0} \\
	 	\hline\hline 
		\end{tabular}
		\hfill
		\begin{tabular}{c|cccccc}
	      \multicolumn{7}{c}{Steven Operator Fit }  \\
	    \hline
		 CEF parameters    & $B_2^0 $ & $B_4^0$ & $B_4^3 $ & $B_6^0  $ & $B_6^3 $ & $B_6^6$  \\
		    (meV)          &  -0.372 & -5.55e-3 & 0.096 & -6.1e-6  & 3.84e-4 & -8.26e-4 \\
		 \hline
		 $E^\textrm{CEF}_i$ (meV) & \multicolumn{6}{c}{ \underline{0},  \underline{0.09},  41.5,   52.3,  \underline{61.6}, \underline{ 78.2},  105.7,  122.8, \underline{126.0}}\\
		\hline
		 $\ket{0}$   & \multicolumn{6}{l}{   $0.653(\ket{6}-\ket{-6})-0.272(\ket{3}+\ket{-3})$}\\
	      $\ket{1}$   & \multicolumn{6}{l}{   $0.656(\ket{6}+\ket{-6})-0.183(\ket{3}-\ket{-3})+ 0.270\ket{0}$}\\
	      \hline
	      $g_\mathrm{eff}$   & \multicolumn{6}{c}{ 12.7} \\
	 	\hline\hline 
		\end{tabular}
\end{table*}

\section{High-field spin excitations}
\label{high-field}
Similar to other rare-earth oxides with large Ising moments, we expect both nearest-neighbor (NN) exchange coupling $J_\mathrm{nn}$ and long-range dipole-dipole interactions $D_{ij}$ in TmMgGaO$_4$. Given the strong quantum fluctuations in the frustrated transverse Ising model, it is non-trivial to determine the value of $J_\mathrm{nn}$ and $\Delta$ based on the behavior of the system in zero field \cite{Li2020nc}. Instead, we apply a strong magnetic field ($\mu_0H$) along the $c$ axis to bring TmMgGaO$_4$ into the spin polarized state \cite{Li2020} and suppress quantum effects~\cite{Ross2011,Paddison2017}. 
\blue{Given Ising Tm$^{3+}$ spins and the multipole nature of the transverse components, the effects of the transverse field are well described by a magnetic field acting along the $x$-axis with the spin excitations properly modeled in the $\mathcal{S}^{zz}$ component of the dynamical structure factor in  linear spin wave theory. } 
Low-energy neutron-scattering measurements were conducted on the MACS spectrometer~\cite{Rodriguez2008} at the NIST Center for Neutron Research with $\mu_0H = 5.5$\,T and $T = 0.1$\,K. These reveal a broad and weakly dispersive spin-wave spectrum [Fig.\ref{Fig:INS}(a)], which resembles that of YbMgGaO$_4$ \cite{Paddison2017, Li2017}. This mode is much broader than the instrumental resolution of $\approx$~0.2 meV and observations in zero field \cite{Shen2019}. For example, the full-width at half-maximum (FWHM) of the excitation at the $M$-point is approximately  0.6 meV [Fig. \ref{Fig:INS}(b)]. Retaining the fitted peak center from fixed $Q$ cuts, we perform fits to the spin wave excitations [black dots in Fig. \ref{Fig:INS}(a)] using the linear spin-wave calculation package SpinW \cite{Toth2015} for the effective spin-1/2 Hamiltonian suitable for TmMgGaO$_4$ \cite{Shen2019, Dun2019, Li2020nc}:
\begin{equation}\label{eq:TIM}
\mathcal{H} =  J_{\mathrm{nn}}\sum_{\langle i,j \rangle}S_i^zS_j^z+ D_{ij}\sum_{i,j}S_i^zS_j^z + \sum_i(\Delta S_i^x+ hg_\mathrm{eff} S_i^z),
\end{equation}
where $\langle i,j \rangle$ indicates nearest neighbor Tm$^{3+}$ pairs, and $D_{ij} =Dr_{\mathrm{nn}}^{3}{\hat{\mathbf{z}}_{i}\cdot\hat{\mathbf{z}}_{j}-3(\hat{\mathbf{z}}_{i}\cdot\hat{\mathbf{r}}_{ij})(\hat{\mathbf{z}}_{j}\cdot\hat{\mathbf{r}}_{ij})}/{r_{ij}^{3}}$ with $r_{\mathrm{nn}}$ equal to the NN Tm-Tm distance and $D = \mu_{0}(g_\mathrm{eff}\mu_\mathrm{B})^2/({4\pi}{k_{\mathrm{B}}r_{\mathrm{nn}}^{3}})= 0.234$\,meV. 
\blue{Therefore,  $D_{ij}$, including intra- and inter-layer couplings, can be directly calculated  and manually added into SpinW. We considered $D_{ij}$ terms up to the 10th nearest neighbor so that the change of energy per spin of the polarized state is less than 0.001\,meV.}
The best fit to the spectrum [red dashed line in Fig. \ref{Fig:INS}(a)] is obtained for
\begin{align}
 J_\mathrm{nn} = 0.557(1)\, \mathrm{meV}, \; \Delta = 0.6(1)\,\mathrm{meV}. \nonumber
 \end{align}
The value of $\Delta$ is in reasonable agreement with previous studies \cite{Li2020nc,Li2020}, while  our obtained $J_\mathrm{nn}$ is about 20\% smaller, mainly due to the inclusion of $D_{ij}$ beyond the 2nd NN.  If we cut off $D_{ij}$ at the 2nd nearest neighbor, our effective spin Hamiltonian is back to the $J1-J2$ model that was used in previous studies \cite{Shen2019, Li2020nc, Li2020},  for which $J2$ is fixed to be 0.045\,meV and our best fit to the dispersion curve [Fig.~3(c) of the main text] gives $J1 = 0.8667(5)$\,meV and $\Delta = 0.66(9)$\, meV. 
We note that the broadening in both the CEF excitations and the high-field spin-wave spectra is likely due to a distribution of $g_\mathrm{eff}$ and $\Delta$ values associated with the Mg-Ga structural disorder \cite{Li2020}. In this sense, the fitted values of $\Delta$ and $J_\mathrm{nn}$ are a local-structure average.

\begin{figure}[tbp!]
	\linespread{1}
	\par
	\begin{center}
		\includegraphics[width=\columnwidth ]{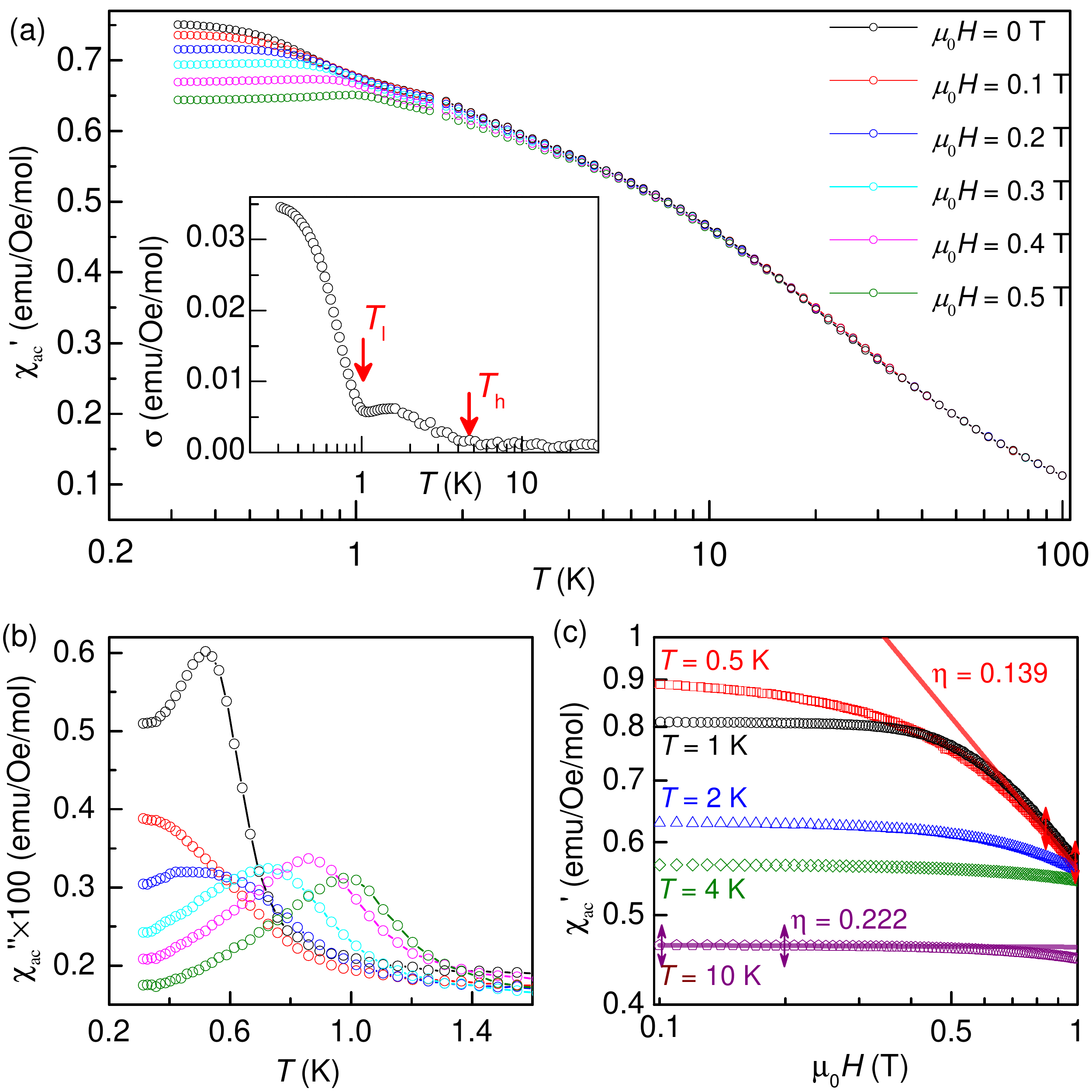}
	\end{center}
	\par
	\caption{\label{Fig:AC}  (a) Real part of the ac susceptibility ($\chi_\mathrm{ac}'$) measured under different dc magnetic  fields $\mu_0H$ appied along the crystallographic $c$-axis and an ac field of 10 Oe with a frequency of 80 Hz. Inset: standard deviation between the six curves. (b) Imaginary part of  ac susceptibility below 1.7\,K  measured under different magnetic dc fields $h$.  (c) Real part of ac susceptibility as a function of $h$ measured at different temperatures in a log-log scale. Selected fits to the power law scaling behavior, $\chi \sim h^{-\frac{4-18\eta}{4-9\eta}}$, are shown as solid lines, and the fitting ranges are indicated by arrows.} 
\end{figure}

\begin{figure*}[tbp!]
	\linespread{1}
	\begin{center}
		\includegraphics[width= 6 in]{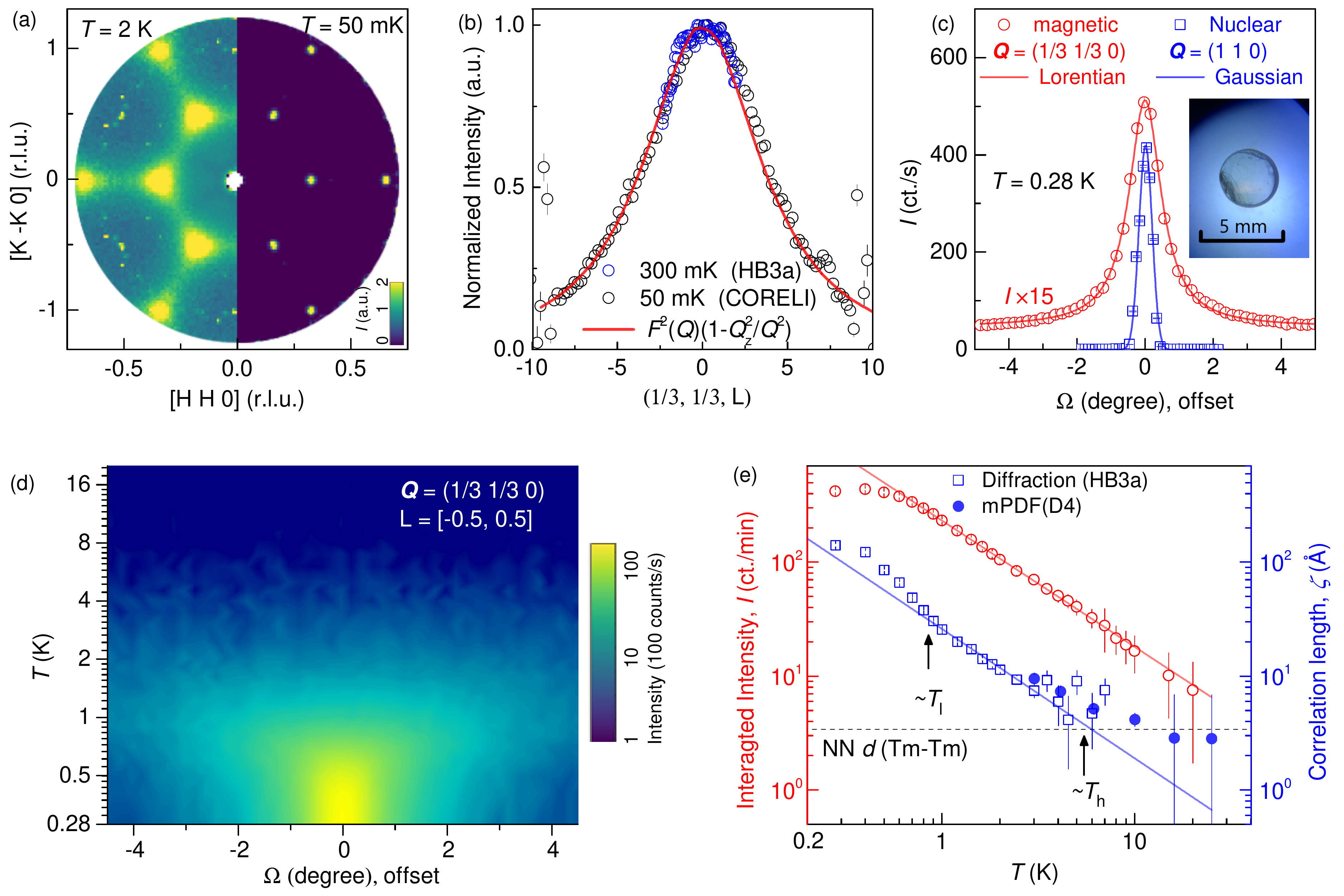}
	\end{center}
	\caption{\label{Fig:Diffraction} (a) Magnetic diffuse scattering of the triangular-lattice Brillouin zone measured on CORELLI at $T=$ 0.05\,K, and 2\,K, where an empty cryostat and a measurement at 40 K were used as background, respectively. Intensities were integrated within $L = \pm0.1$ reciprocal-lattice units (r.l.u.) and symmetrized according to the $-3m$ point group of the Tm site. (b)  $L$-dependence of the $K$-point magnetic scattering measured  on CORRELLI  at 50\,mK (black circle) and  HB3A at 0.28\,K (blue square). The red line represents a calculation based on uncorrelated triangular planes where $F(Q)$ is the neutron magnetic form factor of Tm$^{3+}$. (c) Line shapes of magnetic and nuclear Bragg peaks measured on HB3A at $T$ = 0.28\,K. Inset: A disk-shaped single crystal sample used for the measurement. (d) Temperature dependence of the $K$-point magnetic Bragg peak from rocking-curve $\Omega$-scans. A scan at 40\,K was used for background subtraction. (e) Log-log temperature dependence of the $K$-point peak intensity ($I$, red circle), and 2D correlation length ($\xi$, blue squares and circles) within the triangular layers. Here, $I$ is obtained by integrating $\Omega$ over $[-4.5, 4.5]^{\circ}$. Solid lines represent a linear fit to log-log data points between 1\,K and 5\,K. The dashed line represents the NN Tm-Tm distance.}
\end{figure*}

\section{ AC Susceptibility \& Scaling}
Considering that structural disorder effects are important for the magnetization process of TmMgGaO$_4$ \cite{Li2020} and that the predicted KT phase is fragile against external perturbations \cite{Liu2020}, the question naturally arises as to whether the predicted KT transitions in TmMgGaO$_4$ survives in the zero field limit. To answer this question, we first examine the scaling behavior of the susceptibility in the proposed KT regime, $\chi(H, T) = H^{-\alpha}$ with $\alpha$ = $\frac{4-18\eta (T)}{4-9\eta (T)}$ \cite{Damle2015, Biswas2018}. Here, $\eta (T)$ is the anomalous dimension exponent of the emergent order parameter~\cite{Li2020nc}, which for small $H$ is predicted to be: (i) $\eta (T) = 2/9$ above $T > T_{h}$ so that $\chi$ remains field-independent;  (ii) $\eta (T) \in [2/9, 1/9]$ between $T_{h}$ and $T_{l}$, leading to $H$-dependence of $\chi$ below $T_{h}$ and divergence at $T_{l}$ in the $H \rightarrow 0$ limit; and (iii) $\chi$ becomes flat below  $T_{l}$ with the onset of 3SL ordering. Our ac susceptibility measurements \blue{show a high level of agreement} with these predictions. The $\chi_\mathrm{ac}'(T)$ curves begin to show noticeable $H$-dependence around 5\,K and strongly deviate from each other below \,1 K [Fig.~\ref{Fig:AC} (a)]. To better illustrate this effect, we take the six data sets for $\chi_\mathrm{ac}'(T)$ corresponding to the different applied fields and compute the standard deviation at each temperature [Fig.\ref{Fig:AC} (a) inset]. Clear features are visible around 5~K and 0.9~K, which we tentatively ascribe to the two predicted transitions at $T_h$ and $T_l$ bounding the proposed KT phase in TmMgGaO$_4$. We note, however, that the ac susceptibility data deviate from theoretical predictions in certain ways. First, we do not see a divergence of $\chi_\mathrm{ac}'(T)$ at $T_{l}$ as $H \rightarrow 0$; instead, a peak around 0.6 \,K  appears in the imaginary part which quickly disappears at 0.1\,T, with a broad peak shows up at 0.2\,T and further moves to higher temperature with increasing $H$  [Fig.\ref{Fig:AC} (b)]. Second, and perhaps more significantly, we are unable to obtain a physically meaningful value of $\eta (T)$ from the $\chi_\mathrm{ac}'(H)$ measurements at fixed temperature. \blue{ Instead of a power-law scaling of susceptibility \cite{Damle2015, Biswas2018, Li2020nc},  which appears as a line in a log-log plot, our measured $\chi^{'}_{ac}(h)$ is always a concave function from 0.5\,K to 10\,K [Fig.\ref{Fig:AC} (c)]. Therefore, the fitted value of $\eta$ ranges from 0.139 to 0.222, and strongly depends on the choice of fitting range [some selected fits are shown as dashed lines in Fig.~\ref{Fig:AC} (c)] and $H$. As discussed above, these deviations may be related to the structural disorder; indeed, it was recently proposed that weak, quenched Mg-Ga disorder would drive the emergent 2D KT phase into a gauge glass \cite{huang2020emergent} and thus violate the critical scaling. }

\section{Magnetic Neutron Diffraction}
With evidence for two transitions seen in $\chi_\mathrm{ac}'$, we now turn to diffuse scattering measurements performed on the CORELLI spectrometer~\cite{Ye2018} at ORNL to probe the elastic spin correlations within and below the proposed KT phase. At 0.05\,K, we observe magnetic Bragg peaks at the $K$-points of the triangular Brillouin zone (BZ) [Fig.~\ref{Fig:Diffraction} (a)] corresponding to the 3SL order reported in previous studies \cite{Shen2019,Li2020}.  \blue{The $L$-dependence  of the $K$-point scattering at 50\,mK is shown in Fig.~\ref{Fig:Diffraction} (b), and the intensity is well described by the magnetic form factor [$F^2 (Q)$] of Tm$^{3+}$ multiplied by the polarization factor [1-$Q^2_z/Q^2$]. This means that there are no noticeable inter-layer Tm$^{3+}$-Tm$^{3+}$ correlations down to 50\, mK, and the magnetic behavior of TmMgGaO$_4$ can be described by isolated Tm triangular layers through all temperature regimes in our study.}
At 2\,K (tentatively in the middle of the proposed KT phase), we see that the magnetic Bragg peaks have evolved into broad diffuse scattering with a triangular pattern, indicative of short-range 3SL correlations. These are similar to predictions from Quantum Monte-Carlo~\cite{Li2020nc}, although we do not observe the additional intensity predicted at the $M$-point within the sensitivity of our experiment.

To track the temperature dependence of the $K$-point intensity, we employed the HB-3A diffractometer \cite{Chakoumakos2011} at ORNL with $\lambda\!=\!1.551$~\AA. We find that the magnetic Bragg peaks at 0.28~K display a Lorentzian shape with a FWHM of  $ 1.15(1)^\circ$, which is larger than the instrumental resolution and sample mosaic of $\Omega_G^0 = 0.53(1)^\circ$ obtained through a Gaussian fit to the nuclear Bragg peaks [Fig.~\ref{Fig:Diffraction} (c)]. This points to a finite correlation length within the 2D triangular layers. From the evolution of the $K$-point correlations between 0.28 and 20\,K [Fig.~\ref{Fig:Diffraction} (d)], we extract the integrated peak intensity $I$ and the intra-plane magnetic correlation length $\xi$ according to $\xi = \lambda/(2\Omega_L \sin\theta)$, where $\theta$ is the Bragg scattering angle and $\Omega_L$ the fitted Lorentzian peak width from a Voigt function with fixed $\Omega_G^0$ . Unlike two previous studies which reported a sharp transition at $T_l\approx 1\,K $ based on the temperature dependence of the $K$-point intensity \cite{Shen2019,Li2020}, we see a continuous increase in both $I$ and $\xi$ as the temperature is lowered over a wide range [Fig.~\ref{Fig:Diffraction} (e)], with a clear exponential dependence above 1\,K. Subtle changes are observed around 0.9\,K, evidenced by a flattening of $I$ and an upturn in $\xi$ upon further reduction of the temperature. These observations deviate from a conventional Landau first- or second-order transition, where the magnetic order parameter vanishes above the transition temperature \cite{toledano1987}, and provide further evidence for the gradual evolution of the 3SL order with an extended regime of short-range correlations that could support the KT phase.  While the integrated intensity follows the exponential-law behavior up to 20\,K, the signal becomes extremely broad above 3\,K, limiting the range of trustworthy calculated values for $\xi$ [Fig.~\ref{Fig:Diffraction} (d)(e)]. 

\begin{figure}[tbp!]
	\linespread{1}
	\begin{center}
		\includegraphics[width=\columnwidth  ]{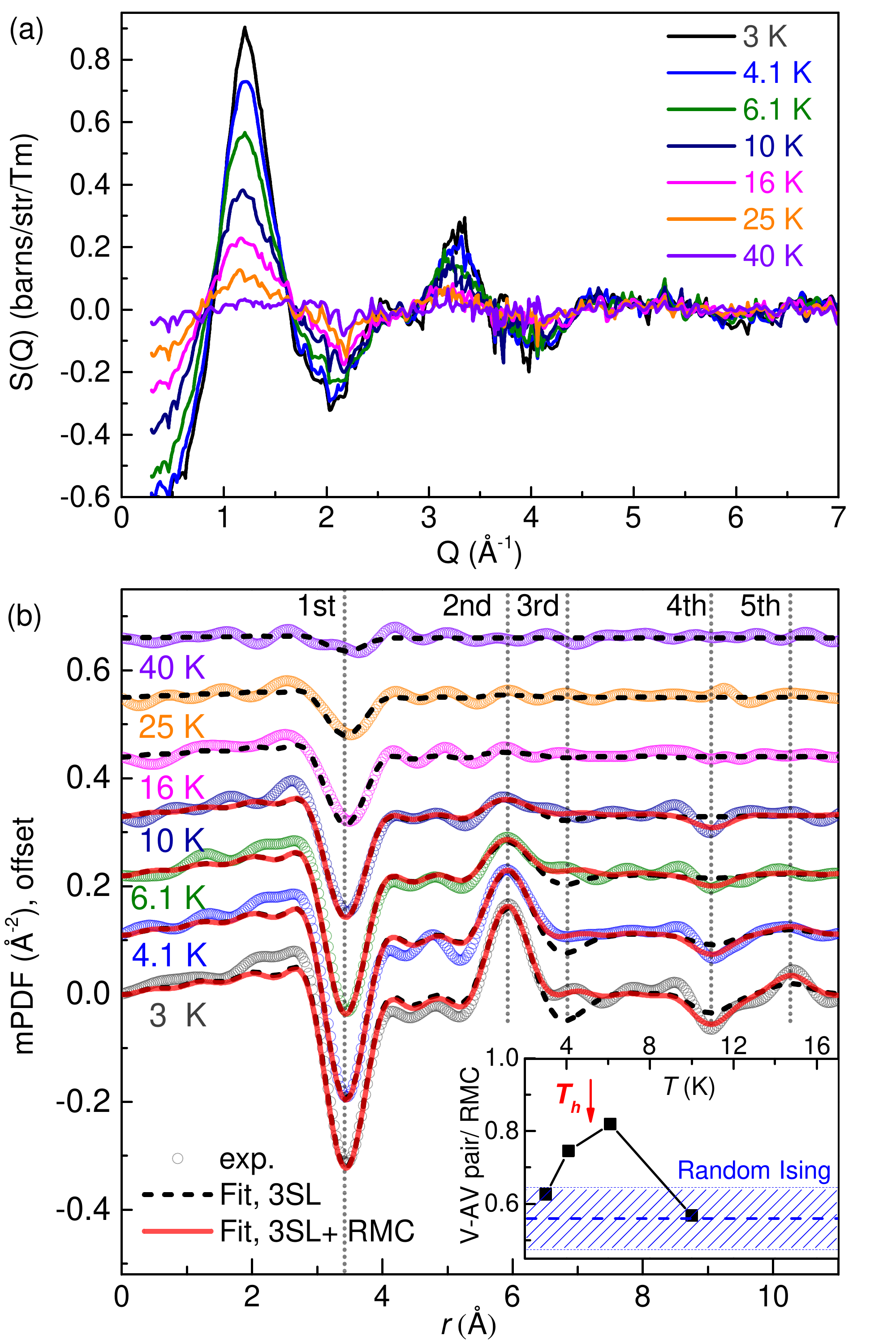}
	\end{center}
	\caption{\label{Fig:mPDF}   
(a) Magnetic scattering cross section of TmMgGaO$_4$  measured on the D4 instrument at ILL at different temperatures. The scattering pattern at 50~K was subtracted from each of the data sets to remove the nuclear and paramagnetic scattering. (b) Magnetic pair distribution function (mPDF) patterns (open circles) obtained at different temperatures. Dashed lines show the best fits to data using a model of the 3SL order. Solid red lines represent alternative fits combining the 3SL model with Reverse Monte Carlo (RMC) simulations of Ising spins. The ${n}$th NN in-plane Tm-Tm distances are illustrated by the vertical dotted lines. Inset: Number of bounded V-AV pairs per RMC configuration as a function of temperature. For comparison, the horizontal blue dashed line shows the average number of V-AV pairs for random Ising configurations, with the shaded area representing one standard deviation.}
\end{figure}

\section{Magnetic Pair Distribution Function}
Further evidence of the proposed KT phase can be gained by examining the spin correlations in real space. We accomplish this through the magnetic pair distribution function (mPDF)~\cite{frand;aca14,frand;aca15}, which is the Fourier transform of the magnetic scattering. 
\blue{The magnetic scattering patterns displayed in Fig.~\ref{Fig:mPDF}(a)  were obtained from energy-integrated measurements of a powder sample on the D4 diffractometer at the Institut Laue-Langevin \cite{D4data, Fischer2002}. The scattering pattern collected at 50~K was used as a reference measurement and was subtracted from all scattering patterns at lower temperature to remove the nuclear and paramagnetic contributions. This is justified because the sample is purely paramagnetic at 50~K (i.e. the correlations between neighboring spins are completely random) and the atomic structure changes very little below 50~K, so only the temperature-dependent part of the total scattering data comes from the magnetic correlations that develop as the temperature is lowered. The broad and strongly temperature-dependent features of the magnetic scattering patterns in Fig.~\ref{Fig:mPDF}(a) demonstrate the development of short-range magnetic correlations with decreasing temperature, as expected. 

The magnetic scattering patterns were then Fourier transformed with $Q_{\mathrm{max}}\!=\!8~\mathrm{\AA^{-1}}$ to generate the real-space magnetic pair distribution function, denoted here as $m$PDF$(r)$ [Fig.~\ref{Fig:mPDF} (b)]. } The negative peak  observed at the NN distance for all temperatures shown in Fig.~\ref{Fig:mPDF}(b) arises from robust NN antiferromagnetic correlations. Additional features in the mPDF data are captured by fits [dashed black lines in Fig.~\ref{Fig:mPDF}] using the reported 3SL model with a finite correlation length $\xi$. The best-fit values for $\xi$ [Fig.~\ref{Fig:Diffraction} (e)] agree well with the  single-crystal diffraction analysis. The small value of $\xi$ above $\approx$10~K indicates that only generic antiferromagnetic correlations between NN spins remain. Additional details about the implementation of the short-range 3SL model are provided in the Supplemental Information \cite{Supplymental}.

Closer inspection of the low-temperature mPDF fits [Fig.~\ref{Fig:mPDF}(b)] reveals small but systematic misfits at the third, fourth, and fifth NN correlations. \blue{These misfits are present for different choices of $Q_{\mathrm{max}}$, indicating they are not simply a consequence of termination error in the Fourier transform \cite{Supplymental}. They are also unlikely to be the result of random noise in the data, since the misfits appear systematically at specific coordination distances. Considering that the misfits are also temperature dependent, we can rule out structural disorder as the origin, since we expect any signatures of structural disorder to be independent of temperature at such low temperatures. We therefore consider this to be a genuine feature of the data meriting further investigation.} 

Interestingly, the third NN distance shows a small, positive feature corresponding to net ferromagnetic correlations at this distance. This is at odds with the antiferromagnetic correlations predicted by the 3SL model, and may instead hint at a small component with stripe-like correlations, which have ferromagnetic alignment between third NN spins~\cite{Li2020nc}. To gain further insight, we performed fits combining the 3SL model with a reverse Monte Carlo (RMC) algorithm~\cite{keen;jpcm91}. \blue{We included two independent components in our modeling: the mPDF from short-range 3SL correlations and the mPDF from a sheet of Ising spins optimized through random flipping by the RMC algorithm. The Ising spins had constant magnitudes and were fixed at the positions of the Tm atoms in a circle of diameter 50~\AA\ in the \textit{ab} plane. At each iteration of the algorithm, we performed a least-squares optimization using three independent parameters: a scale factor for the Ising spin component, a scale factor for the 3SL component, and the correlation length for the 3SL component. The algorithm ran until convergence at the level of $\sim$2\% was reached, which typically occurred within a few hundred iterations. 110 RMC refinements were done for each temperature, producing a narrow distribution of values for the goodness of fit, the two scale factors, and the 3SL correlation length, with negligible correlation between the parameters \cite{Supplymental}. This boosts our confidence that the RMC algorithm converged reliably and that the resulting spin configurations are meaningful.}  Representative fits produced from this combined 3SL+RMC fitting approach are shown by the solid red curves in Fig.~\ref{Fig:mPDF}(b) for 3, 4.1, 6.1, and 10~K. These fits clearly correct the mismatches left by the 3SL model, indicating that the RMC-produced Ising spin configurations should be representative of the types of correlations present in the system beyond just the 3SL correlations.

We now look for evidence of the formation of V-AV pairs by inspecting the spin configurations produced by the RMC algorithm [see e.g. Fig.~\ref{Fig:CEF}(b)]. A home-built python script counted the number of V-AV pairs in each configuration, allowing us to find the average number of V-AV pairs formed per RMC configuration among the 110 configurations for each temperature. We designate this quantity as \navg. \blue{A vortex and antivortex were considered to form a bound pair if they shared any constituent pseudospins. For an Ising configuration on the triangular lattice, only two types of bound pairs are possible under this definition, both of which are shown in Fig.~\ref{Fig:CEF}(b). To determine whether the RMC fits provide evidence for preferential V-AV formation around the expected KT transition temperature, we performed a hypothesis test for the difference in means, where the null hypothesis is that the RMC fits performed on the data \textit{do not} show any preference for V-AV pair formation beyond what would be expected from completely random Ising spin configurations. The test statistic of interest is then \navg. To implement this hypothesis test, we generated 1000 distinct sets of 110 random Ising spin configurations of the same size as the RMC-generated configurations. We calculated \navg\ for each set of 110 random configurations, and from the resulting 1000 values of \navg, we constructed its underlying null distribution. This distribution is highly Gaussian \cite{Supplymental}, with a mean of 0.56 V-AV pairs per configuration and a standard deviation of 0.08. We can now compare the \navg\ value determined from the RMC fits at each temperature with the null distribution, allowing us to conclude whether or not the RMC fits show a preference for V-AV pair formation. } 

Fig.~\ref{Fig:mPDF} Inset shows the \navg\ calculated from the RMC fits at 3, 4.1, 6.1, and 10~K as the black squares. The mean of the null distribution is given by the horizontal dashed line, and the shaded region represents one standard deviation above and below the mean. The value of \navg\ at 10~K is very close to the mean of the null distribution, indicating no preference for V-AV pair formation. At 6.1~K and 4.1~K, however, \navg\ increases sharply and falls well outside the expectation for random spins given by the null distribution. At these two temperatures, we can reject the null hypothesis (and therefore accept the alternative hypothesis that the RMC fits preferentially result in the formation of V-AV pairs) with high degrees of certainty corresponding to statistical \textit{p}-values of 0.001 and 0.015, respectively. This demonstrates a clear tendency for V-AV pair formation at temperatures in the immediate vicinity of the proposed upper KT transition, precisely where the proliferation of vortices would be expected. As the temperature is lowered further to 3~K, \navg\ decreases again, falling to approximately one standard deviation above the mean of the null distribution. In this case, we can reject the null hypothesis with considerably less confidence ($p = 0.24$), though it is still generally consistent with the preferential formation of V-AV pairs at this temperature. It is also consistent with the notion that as the temperature decreases toward the lower KT transition, the 3SL order becomes increasingly dominant at the expense of the KT correlations giving rise to the V-AV pairs. In any case, these findings give strong support to the formation of V-AV pairs around 4--6~K and should be considered as evidence for the proposed upper KT transition in TmMgGaO$_4$.

\section{ Summary}
In summary, the magnetometry and neutron scattering results presented here establish TmMgGaO$_4$ as a strong candidate for a solid-state system realizing KT physics. Our inelastic neutron scattering measurements confirm the transverse-field Ising model on the triangular lattice as the foundation to understand the magnetism in TmMgGaO$_4$ and help clarify the role played by structural disorder. Magnetometry reveals two transitions around 0.9~K and 5~K, consistent with the theoretical predictions for a KT phase bounded by these transitions. Elastic and energy-integrated neutron scattering measurements confirm the presence of 3SL correlations in the ground state, which become gradually weaker and shorter-range as the temperature is raised. Investigation of the spin correlations in the proposed KT phase in real space via mPDF analysis suggests a tendency to form bound vortex-antivortex pairs around 5~K, which is the hallmark of the proposed KT transition. Structural disorder does not appear to play a dominant role in the zero-field physics of TmMgGaO$_4$, in contrast to YbMgGaO$_4$ \cite{Paddison2017, Li2017,Zhu2017, Kimchi2018}. Our work motivates future studies on the interplay between structural disorder effects and quantum magnetism, while also highlighting the value of mPDF analysis of short-range spin correlations in real space.

\begin{acknowledgements}
The work of Z.L.D, M.D. and M.M. (AC susceptibility, diffraction and inelastic neutron scattering experiments and their analysis) was supported by the National Science Foundation through Grant No. NSF-DMR- 1750186. The work of R.B and B.A.F. was supported by U.S. Department of Energy, Office of Science, Basic Energy Sci- ences through Award No. DE-SC0021134 (magnetic pair-distribution function experiments and reverse Monte-Carlo analysis) . The work of Q.H. and H.Z. was supported by the National Science Foundation through Grant No. NSF-DMR- 2003117 (sample growth and low-temperature AC susceptibility). This research used resources at the High Flux Isotope Reactor and Spallation Neutron Source, a DOE Office of Science User Facility operated by the Oak Ridge National Laboratory. We thank the Institut Laue-Langevin for use of its neutron instrumentation. Access to MACS was provided by the Center for High Resolution Neutron Scattering, a partnership between the National Institute of Standards and Technology and the National Science Foundation under Agreement No. DMR-1508249
\end{acknowledgements}

\bibliographystyle{apsrev4-2}
\bibliography{TMGO}

\end{document}